\documentclass[10pt,aps,prl,twocolumn,showpacs,amsmath,amssymb]{revtex4-2}
\usepackage{hyperref}
\begin{document}

\title{Constraints on Bulk Fields: No-Go Conjectures for Braneworld Models}

\author{G. Alencar}
\email{geova@fisica.ufc.br}
\affiliation{Departamento de F\'{\i}sica, Universidade Federal do Cear\'a, 60451-970 Fortaleza, Cear\'a, Brazil}

\author{R. N. Costa Filho}
\email{rai@fisica.ufc.br}
\affiliation{Departamento de F\'{\i}sica, Universidade Federal do Cear\'a, 60451-970 Fortaleza, Cear\'a, Brazil}

\author{M. Gogberashvili}
\email{gogber@gmail.com}
\affiliation{Ivane Javakhishvili Tbilisi State University, 3 Chavchavadze Avenue, Tbilisi 0179, Georgia}

\begin{abstract}

This work establishes a series of no-go conjectures that impose rigorous constraints on the zero mode localization of bulk fields in braneworld scenarios, specifically affecting gauge and spinor fields within five-dimensional spacetimes. Our approach differs from traditional methods as it does not rely on specific equations of motion, making our results broadly applicable across various braneworld models. These no-go conditions reveal fundamental limitations in field localization, challenging the feasibility of embedding fields on the brane. For instance, our analysis demonstrates that existing models fail to achieve consistent localization for gauge and spinor fields. Additionally, one of our conditions indicates that the effective Lagrangian on the brane cannot exhibit conformal invariance.

\end{abstract}


\maketitle


Braneworld models have become a key framework in modern theoretical physics, offering promising solutions to fundamental problems such as the hierarchy problem and the unification of forces \cite{Gogberashvili:1998vx, Gogberashvili:1999tb, Randall:1999ee, Randall:1999vf}. In these models, our four-dimensional universe is embedded within a higher-dimensional space, where additional spatial dimensions help explain gravitational interactions on vastly different scales \cite{Maartens:2010ar, Raych}.

In the RS framework, two main approaches are commonly used to incorporate matter fields. The first approach assumes that the stress-energy tensor is strictly confined to the brane. In this case, the effective gravitational equations on the brane are obtained by projecting the bulk Einstein equations using the Gauss-Codazzi formalism. This method, developed by Shiromizu, Maeda, and Sasaki \cite{Shiromizu:1999wj}, results in a four-dimensional effective theory where the influence of the bulk is encoded in additional terms, such as the Weyl contribution, which captures gravitational effects from the extra dimension. While this approach provides a direct route to obtaining an effective four-dimensional description of gravity, it does not address whether fields can naturally be confined to the brane.

A more fundamental approach is considering that matter fields exist and live in the bulk. In this framework, one first derives the equations of motion in the higher-dimensional spacetime and then analyzes whether the zero-modes of fields can be dynamically confined to the brane. The standard criterion for localization is that, after solving the equations of motion for the zero mode of a given matter field, the integral of the action over the extra dimensions remains finite. This perspective is essential for understanding the zero-mode localization mechanisms of fundamental fields, such as gauge fields and fermions, and plays a key role in constructing consistent higher-dimensional models. However, in the presence of gravity, achieving exact localization of zero modes is not always possible \cite{Fichet:2019owx}. Even when zero mode localization is achieved, if one considers the whole interplay between gravitational and field dynamics, a finite integral over the extra dimensions alone is insufficient to ensure a consistent theory.

This idea was first explored by Duff et al. in the context of p-form fields in Type II braneworlds. They demonstrated that ensuring consistency requires more than just the finiteness of the action integral over the extra dimensions—it also demands consistency with the full system of Einstein equations, taking into account the backreaction of localized fields on the geometry. They applied this gravitational consistency condition to p-form fields and showed that only the 0-form and its dual satisfy the full Einstein equations \cite{Duff:2000se}. This result effectively rules out other p-form fields, even though the authors of Ref. \cite{Kaloper:2000xa} had previously shown that their corresponding integrals over the extra dimensions converge. This idea was later expanded upon by several authors, who developed additional consistency conditions to constrain further the dynamics of bulk fields and their zero mode localization properties \cite{Gibbons:2000tf, Leblond:2001xr, Freitas:2020mxr}. These conditions determine which fields can be consistently confined to the brane while ensuring a self-consistent gravitational and field-theoretical framework.

In this work, we build on the ideas of Duff et al. and investigate the obstructions that arise from the interplay between gravity and general matter fields in the RS framework.


Consider the brane metric in a 5D bulk spacetime with the signature $(-,+,+,+,+)$,
\begin{equation} \label{ansatz}
ds^2 = e^{2 A(y)} g_{\alpha\beta} \left(x^\gamma\right) dx^\alpha dx^\beta + dy^2~,
\end{equation}
where Greek letters numerate 4D coordinates ($\alpha, \beta, \gamma = 0, 1, 2, 3$), $y$ denotes the space-like extra dimension and $A(y)$ is a warp function. For the metric \textit{ansatz} \eqref{ansatz}, the components of 5D Einstein tensor and the fifth component of the Ricci tensor yield:
\begin{align}
{}^{(5)}G_{\mu\nu} =& G_{\mu\nu} + 3\left(A'' + 2A'^2\right)e^{2A}g_{\mu\nu}~,\label{Gmunu}\\
{}^{(5)}G_{5\nu} =&0~,\label{G5nu}\\
{}^{(5)}G_{55} =&-\frac{1}{2} e^{-2 A}R +6 A'^2~, \label{G55} \\
{}^{(5)}R_{55} =& - 4\left(A'' + A'^2\right)~,\label{R55}
\end{align}
where primes stand for derivatives with respect to $y$, and $G_{\mu\nu}$ and $R$ are the 4D Einstein tensor and the Ricci scalar, respectively.

To derive an effective gravitational field equation in four dimensions that include contributions from the extra dimensions, we multiply \eqref{Gmunu} by the factor $-(1-n)e^{(n-2)A}$ (where $n$ is a constant) and add it to \eqref{R55} multiplied by $3(n-2)g_{\mu\nu}/4$. This yields:
\begin{equation}
\begin{split} \label{sum}
(1-n) G_{\mu\nu}  &= (1-n){}^{(5)}G_{\mu\nu}- \frac{3(n-2)}{4} g_{\mu\nu} {}^{(5)}R_{55} +\\
&+ 3 \left(A'e^{nA}\right)' e^{-nA}g_{\mu\nu}~.
\end{split}
\end{equation}
Using the 5D Einstein equations,
\begin{equation}
{}^{(5)}G_{\mu\nu} = 8\pi \mathbf{G} T_{\mu\nu}~,
\end{equation}
where $\mathbf{G}$ denotes the bulk gravitational constant, the relation \eqref{sum} can be expressed as:
\begin{equation}
\begin{split} \label{tensor sum}
(1-n)G_{\mu\nu} &= 8\pi \mathbf{G}(1-n) T_{\mu\nu} +\\
&+ 2\pi (n-2)\mathbf{G}e^{2A}\left(T_{\alpha}{}^{\alpha} - 2T_5{}^5\right) g_{\mu\nu} +\\
&+ 3\left(A'' + nA'^2\right)e^{2A} g_{\mu\nu} ~.
\end{split}
\end{equation}
These tensorial conditions impose stricter constraints than the well-known scalar consistency condition \cite{Gibbons:2000tf, Leblond:2001xr}. They can assist in identifying braneworlds with properly localized modes without the need to solve the equations of motion for specific matter fields. The scalar consistency condition \cite{Gibbons:2000tf, Leblond:2001xr}
\begin{equation}
\begin{split} \label{Gibbons}
\left(A'e^{nA}\right)' &= \frac{2\pi \mathbf{G}}{3}e^{nA}\left[T_{\mu}{}^{\mu} + 2(n-2)T_{5}{}^{5}\right] - \\
&- \frac{1-n}{12}e^{(n-2)A}R~,
\end{split}
\end{equation}
is a particular case of \eqref{tensor sum}. If we consider a compact internal space without boundary, specifically the $S^1/Z_2$ orbifold $y \in  [- y_\pi , y_\pi]$, and a smooth warp function $A(y)$, the integral of the left-hand side of \eqref{Gibbons} vanishes, leading to the known consistency relation \cite{Gibbons:2000tf, Leblond:2001xr}:
\begin{equation}\label{GSR}
\oint e^{nA}\left[T_{\mu}{}^{\mu} + 2(n-2)T_{5}{}^{5}\right] = \frac{1-n}{8\pi \mathbf{G}}\oint e^{(n-2)A}R~.
\end{equation}
The scalar relation \eqref{GSR} is derived by integrating \eqref{Gibbons} over the extra dimension $y$, which effectively averages the expression and introduces certain limitations. First, the condition in \eqref{GSR} holds only for RS I-type braneworlds, where the integral of the last term in \eqref{tensor sum} (or the left-hand side of \eqref{Gibbons}) vanishes, as the integrals of the two delta functions cancel each other. Second, integration over $y$ may lead to the loss of certain local properties, which could impose additional constraints.

Before deriving the conditions on bulk fields, let us obtain general results from the expression \eqref{tensor sum}. For $n = 1$ we have
\begin{equation}\label{localn1}
A'' + A'^2 = \frac{2\pi \mathbf{G}}{3} \left(T_{\mu}{}^{\mu} - 2T_5{}^5\right) ~.
\end{equation}
Inserting this back into \eqref{tensor sum} and evaluating it for $n = 0$, we obtain:
\begin{equation}
\begin{split} \label{EEBulk}
G_{\mu\nu} &= 8\pi \mathbf{G} T_{\mu\nu} - 2\pi \mathbf{G} e^{2A} \left(T_{\alpha}{}^{\alpha} - 2T_5{}^5\right) g_{\mu\nu} +\\
&+ 3A'^2 e^{2A} g_{\mu\nu}~.
\end{split}
\end{equation}
We can derive the background solution for the RS models from the above two equations. In general, we must consider a vacuum source, denoted as ${}^{(v)}T_{MN}(y)$, with capital Latin indices ($M, N, \ldots$) representing the coordinates of the 5D spacetime. This source can describe one or two delta branes or the source for smooth-type RS II models. In all these cases, the solutions are obtained from:
\begin{eqnarray}
\Lambda_b g_{\mu\nu} &=& 8\pi \mathbf{G} T^{(v)}_{\mu\nu}(y) - 3A'^2 e^{2A} g_{\mu\nu} - \\
- &2\pi& \mathbf{G} e^{2A} \left[{}^{(v)}T_{\alpha}{}^{\alpha}(y) - 2{}^{(v)}T_5{}^5(y)\right] g_{\mu\nu}~,\label{EEBulkVmunu} \\
A'' + A'^2 &=& \frac{2\pi \mathbf{G}}{3} \left[{}^{(v)}T_{\mu}{}^{\mu} - 2{}^{(v)}T_5{}^5\right] ~,\label{localn1V}
\end{eqnarray}
where
\begin{equation} \label{local2}
\Lambda_b = 4\pi \mathbf{G}e^{2A}T_5{}^5 + 3 A'^2 e^{2A}
\end{equation}
serves as the brane cosmological constant. It can be verified that the flat solution of both RS models satisfies this condition with $\Lambda_b = 0$.

Finally, we consider adding a bulk field to the action, in addition to ${}^{(v)}T_{MN}(y)$, such that:
\begin{equation}\label{sumvacumm}
T_{MN} = {}^{(v)}T_{MN}(y) + {}^{(b)}T_{MN}(x,y)
\end{equation}
where '$(b)$' means bulk field. From equations \eqref{localn1}, \eqref{EEBulk}, \eqref{EEBulkVmunu}, \eqref{localn1V}, and \eqref{sumvacumm}, we can derive four consistency conditions for the bulk energy-momentum tensor ${}^{(b)}T_{MN} \left( x^\alpha, y \right)$ alone, without relying on the equations of motion of matter fields, thereby generalizing the result from \cite{Freitas:2020mxr}."

Let us present these conditions; the second and fourth are new and have not been considered previously.

\begin{enumerate}

\item One restriction on the bulk energy-momentum tensor follows directly from \eqref{G5nu}:
\begin{equation} \label{cond01}
^{(b)}T_{\mu5} \left( x^\alpha, y \right) = 0~.
\end{equation}

\item To obtain an additional condition, we substitute \eqref{localn1V} and \eqref{sumvacumm} into \eqref{localn1}, yielding the following constraint:
\begin{equation} \label{cond03}
    {}^{(b)}T_{\mu}{}^{\mu} - 2{}^{(b)}T_5{}^5 = 0~.
\end{equation}
This condition introduces a \textbf{new} consistency requirement that limits many models satisfying \eqref{cond01} \cite{Freitas:2020mxr}. Furthermore, \eqref{cond03} implies that unless $T_5{}^5 = 0$, the effective action on the brane cannot be conformally invariant.

\item Another condition can be derived by substituting \eqref{EEBulkVmunu}, \eqref{sumvacumm}, and \eqref{cond03} into \eqref{EEBulk}, yielding the following equation:
\begin{equation}\label{EEBulkeff}
G_{\mu\nu} = 8\pi \mathbf{G} {}^{(b)}T_{\mu\nu} + \Lambda g_{\mu\nu}~.
\end{equation}
Since the left-hand side of \eqref{EEBulkeff} depends only on the 4D coordinates $x^\alpha$, the right-hand side must also be independent of $y$. This leads to the conditions:
\begin{equation}\label{cond02}
{}^{(b)}T_{\mu\nu}\left( x^\alpha, y \right) = {}^{(b)}T_{\mu\nu}\left( x^\alpha \right)~,
\end{equation}
which was first identified in \cite{Duff:2000se} and applied to $p$-form fields.

\item The final consistency relation arises from \eqref{G55}, which requires that the extra-dimensional component of the energy-momentum tensor has the following structure:
\begin{equation} \label{cond04}
{}^{(b)}T_{55}\left( x^\alpha, y \right) = -\frac{1}{16\pi \mathbf{G}} e^{-2 A} f\left( x^\alpha \right)~,
\end{equation}
where $f\left( x^\alpha \right)$ is a function that depends only on the brane coordinates. This is also a \textbf{new} condition.

\end{enumerate}
Let us now explore some specific cases to understand these implications better. From this point onward, we will omit the '$(b)$' notation for bulk fields.


{\bf Scalar Fields:} Consider the scalar field $\Phi$, with the following energy-momentum tensor:
\begin{equation} \label{Tscalar}
\begin{split}
T_{5\mu} &= \partial_5 \Phi\, \partial_\mu \Phi~, \\
T_{\mu\nu} &= \partial_\mu \Phi\, \partial_\nu \Phi - \\
&\quad - \frac{1}{2} g_{\mu\nu} e^{2A} \left[ \Phi'^2 + e^{-2A} \partial_\mu \Phi \, \partial^\mu \Phi + V(\Phi) \right]~, \\
T_{\mu}{}^\mu &- T_5{}^5 = -\frac{3}{2} e^{-2A} \Phi'^2 - V(\Phi)~, \\
T_{55} &= \frac{1}{2} \Phi'^2 - \frac{1}{2} e^{-2A} g^{\alpha\beta} \partial_\alpha \Phi \, \partial_\beta \Phi - \frac{1}{2} V(\Phi)~.
\end{split}
\end{equation}
By separating variables as $\Phi (x^N) = \xi_\Phi (y) \phi(x^\nu)$, condition \eqref{cond01} implies that
\begin{equation} \label{xi}
\xi_\Phi(y) \sim \text{constant}~,
\end{equation}
which is consistent with condition \eqref{cond02}. However, condition \eqref{cond03} takes the form
\begin{equation} \label{Phi}
T_{\mu}{}^\mu - 2 T_5{}^5 = -V(\Phi) = 0~,
\end{equation}
which holds only if the potential $V(\Phi) = 0$. Therefore, without relying on the equations of motion, we conclude that only free scalar fields can be consistently localized.


{\bf Fermion Fields:} Consider the 5D fermions $\Psi \left(x^\nu, y\right)$, with the energy-momentum tensor given by:
\begin{equation}
    T_{MN} = \frac{i}{2} \bar{\Psi} \Gamma_{(M} \nabla_{N)} \Psi - \frac{i}{2} \nabla_{(N} \bar{\Psi} \Gamma_{M)} \Psi~.
\end{equation}
By separating variables as $\Psi = \xi(y) \psi\left(x^\nu\right)$, and using the expressions
\begin{equation}
\omega_{\mu}\left(x^\nu, y\right) = \hat{\omega}_{\mu}\left(x^\nu\right) + \frac{1}{2} \Gamma_{\mu} \Gamma^{y} A'~, \quad \omega_{y}\left(x^\nu, y\right) = 0~,
\end{equation}
along with the 5D gamma matrices defined by
\begin{equation}
\Gamma^{\mu} = e^{-A} \gamma^{\mu}~, \quad \Gamma^{5} = -i \gamma^5~,
\end{equation}
we can write the components of the energy-momentum tensor as:
\begin{align}
T_{5\mu} &= \frac{i}{2} \xi^2 \left[\bar{\psi} \gamma_{5} \hat{\nabla}_{\mu} \psi - \hat{\nabla}_{\mu} \bar{\psi} \gamma_{5} \psi \right] - \nonumber\\
&\quad - \frac{i}{2} e^{-A} A' \xi^2 \bar{\psi} \gamma_{\mu} \psi~, \label{Tfermion1}\\
T_{\mu\nu} &= \frac{i}{2} e^{-A} \xi^2 \left[ \bar{\psi} \gamma_{(\mu} \hat{\nabla}_{\nu)} \psi - \hat{\nabla}_{(\mu} \bar{\psi} \gamma_{\nu)} \psi\right] -\nonumber\\
&\quad - 2 i e^{2A} A' \xi^2 g_{\mu\nu} \bar{\psi} \psi~, \label{Tfermion2}\\
T_{\mu}{}^{\mu} &- T_{5}{}^5 = i \xi^2 e^{-3A} \left[\bar{\psi} \gamma^{\mu} \hat{\nabla}_{\mu} \psi - \hat{\nabla}_{\mu} \bar{\psi} \gamma^{\mu}\psi \right] -\nonumber\\
&\quad - 8 i A' \xi^2 \bar{\psi} \psi~, \label{Tfermion3}\\
T_{55} &= 0~. \label{Tfermion4}
\end{align}
From condition \eqref{cond02}, we find that the first two terms in \eqref{Tfermion2} are independent of $y$ if $\xi(y) \sim e^{A/2}$. However, the remaining terms are not independent of $y$, indicating that zero mode localization of both fermion chiralities is impossible. This result suggests that free bulk fermion fields cannot be localized in this simplest case.

One might consider introducing Yukawa couplings of the form $\lambda f(y) \bar{\Psi} \Psi$ to localize fermion fields. However, the only viable function is $f(y) = A'(y)$ \cite{Freitas:2020mxr}, which imposes significant restrictions on potential models. Under this setup, we add the following contribution to the energy-momentum tensor:
\begin{equation}
T_{MN} = -\frac{1}{2} g_{MN} \lambda f(y) \bar{\Psi} \Psi~.
\end{equation}
At first glance, this expression appears favorable, as it satisfies condition \eqref{cond02} if $f = A'$. However, condition \eqref{cond04} requires that:
\begin{equation}
T_{55} = -\frac{1}{2} \lambda f(y) \bar{\Psi} \Psi = -\frac{1}{2} \lambda A' e^{A} \bar{\psi} \psi~,
\end{equation}
which remains consistent only if $\lambda = 0$. Consequently, all models utilizing Yukawa couplings must be excluded.


{\bf Gauge Fields:} We now explore the implications of our consistency conditions for gauge fields, which have an energy-momentum tensor given by:
\begin{equation}
    T_{MN} = F_{MQ}F_{N}{}^Q - \frac 14 g_{MN} F^{PQ}F_{PQ}~,
\end{equation}
where $F_{MN} = \partial_M B_N - \partial_N B_M$ is the field strength tensor. The components of this tensor can be expressed as:
\begin{equation} \label{Tmunuone}
\begin{split}
T_{5\mu} &= F_{5\nu}F_{\mu}{}^\nu~,\\
T_{\mu\nu} &= F_{\mu Q}F_{\nu}{}^Q - \frac 14 g_{\mu\nu} F^{PQ}F_{PQ}~,\\
T_{\mu}{}^{\mu} &- 2T_5{}^5 = -\frac{3}{2} e^{-2A} F^{\mu 5} F_{\mu 5} + \frac{1}{2} e^{-4A} F^{\mu\nu} F_{\mu\nu}~,\\
T_{55} &= \frac{1}{2} F_{5\mu} F_{5}{}^{\mu} - \frac{1}{4} F^{\mu\nu} F_{\mu\nu}~.
\end{split}
\end{equation}
By separating variables for the 5D gauge field as $B_\mu = \xi_B(y) \mathbb{B}_\mu(x^\nu)$ and choosing the gauge $B_5 = 0$, we find that condition \eqref{cond01} implies $\xi_B(y) = \text{constant}$. However, this leads to $T_{\mu\nu} = e^{-2A} F_{\mu\alpha} F_{\nu}{}^{\alpha}$, which is inconsistent with condition \eqref{cond02}. Finally, from conditions \eqref{cond03} and \eqref{cond04}, we obtain:
\begin{align}
    T_{\mu}{}^{\nu} - 2T_{5}{}^{5} &= \frac{1}{2} e^{-4A} F^{\mu\nu} F_{\mu\nu} = 0~, \label{oneformtrace} \\
    T_{55} &= \frac{1}{4} e^{-4A} F^{\mu\nu} F_{\mu\nu}~. \label{oneform55}
\end{align}
This enforces the action for the gauge field to be zero. Therefore, without relying on the equations of motion, we confirm the well-known result that gauge fields cannot be localized. This conclusion is local and applies to both RS models, consistent with the findings of \cite{Davoudiasl:1999tf} in the compact case.

Many zero-mode localization mechanisms for gauge fields involve introducing a coupling term of the form $G(y) F^{PQ}F_{PQ}$ \cite{Kehagias:2000au, Fu:2011pu, Chumbes:2011zt, Landim:2011ki, Landim:2011ts, Alencar:2012en}. Applying the consistency conditions discussed in \cite{Duff:2000se} reveals that $G(y) = e^{2A}$ \cite{Freitas:2020mxr}. In this case, we have the following components of the energy-momentum tensor:
\begin{equation} \label{TmunuoneG}
\begin{split}
T_{5\mu} &= G(y) F_{5\nu} F_{\mu}{}^\nu~,\\
T_{\mu\nu} &= G(y) F_{\mu Q} F_{\nu}{}^Q - \frac{1}{4} g_{\mu\nu} F^{PQ}F_{PQ}~,\\
T_{\mu}{}^{\mu} &- 2 T_5{}^5 = -\frac{3}{2} e^{-2A} G(y) F^{\mu 5} F_{\mu 5} + \\
&+ \frac{1}{2} e^{-4A} G(y) F^{\mu\nu} F_{\mu\nu}~,\\
T_{55} &= G(y) \frac{1}{2} F_{5\mu} F_{5}{}^{\mu} - G(y) \frac{1}{4} F^{\mu\nu} F_{\mu\nu}~.
\end{split}
\end{equation}
The first condition \eqref{cond01} again implies that $\xi_B(y) = \text{constant}$. With this, the second condition \eqref{cond02} is satisfied for $G(y) = e^{2A}$. Consequently, our third condition \eqref{cond03} reduces to:
\begin{equation}
    T_{\mu}{}^{\mu} - 2 T_{5}{}^{5} = \frac{1}{2} e^{-4A} G(y) F^{\mu\nu} F_{\mu\nu} = 0 ~, \label{oneformtraceG}
\end{equation}
which demonstrates that this zero-mode localization mechanism for gauge fields is incompatible with our conditions and must, therefore, be excluded.

One approach to localizing gauge fields is through couplings with the Ricci tensor and Ricci scalar \cite{Alencar:2014moa}, which leads us to consider the 5D energy-momentum tensor:
\begin{equation}
\begin{split} \label{tensor}
T_{MN} = & -g_{MN} \mathcal{L}_{B} + F^Q_M F_{NQ} + \lambda_1 R_{MN} B_P B^P +\\
+ &\lambda_1 R\, B_M B_N + \lambda_1 g_{MN} \Box(B_Q B^Q) -\\
- &\lambda_1 \nabla_M \nabla_N \left( B_Q B^Q \right) + 2 \lambda_2 R_{NQ} B_M B^Q -\\
- &\lambda_2 \nabla_P \nabla_N \left( B^P B_M \right) + \lambda_2 \frac{1}{2} \Box \left( B_M B_N \right) +\\
+ &\lambda_2 \frac{1}{2} g_{MN} \nabla_P \nabla_Q \left( B^P B^Q \right) ~. 
\end{split}
\end{equation}
For simplicity, we consider the case where $B_5 = 0$. It is unnecessary to compute all components of the energy-momentum tensor to identify an inconsistency within the model. The 4D part of the energy-momentum tensor takes the form
\begin{equation}
  T_{\mu\nu} = e^{-2A} \psi^2 F_{\mu \alpha} F_{\nu}{}^\alpha + \ldots ~,
\end{equation}
and, from condition \eqref{cond02}, we obtain $\psi = e^A$. Under this condition, our first constraint \eqref{cond01} reduces to:
\begin{equation}
  T_{5\nu} = e^{-2A} \psi \psi' B_5 F_{\nu}{}^\alpha = 0 ~,
\end{equation}
which is incompatible with the second condition \eqref{cond02}. This incompatibility implies that any zero-mode localization mechanism relying on couplings with gravity must also be ruled out.


In conclusion, this letter addresses the fundamental challenges and constraints associated with the zero mode localization of fields within braneworld scenarios. We derived a set of consistency relations that impose four critical conditions on braneworld models involving quasi-localized (extra-dimension-dependent) bulk fields. By applying these constraints to massless spin-0, spin-1/2, and spin-1 fields, we demonstrated that the implications of the Einstein equations effectively rule out all previously proposed zero mode localization mechanisms for gauge and spinor fields on the brane.

A particularly noteworthy result is that one of our conditions requires the energy-momentum tensor of fields localized on the brane to satisfy $T_\mu{}^\mu = 2 T_5{}^5$. This implies that unless $T_5{}^5 = 0$, the effective action on the brane cannot preserve conformal invariance.

The constraints derived in this paper, which hold independently of the specific equations of motion, offer valuable insights for constructing more realistic braneworld models and advancing our understanding of physics beyond the standard framework. Future research should extend these consistency conditions to higher-dimensional scenarios and investigate their implications for models with interacting bulk fields. An intriguing possibility is that the strong constraints obtained here arise from the ansatz (\ref{ansatz}), which is commonly employed in studies considering bulk fields. If this is the case, revisiting the underlying assumptions of this ansatz could lead to new perspectives on localization mechanisms and the structure of extra-dimensional theories. These directions will open new avenues for exploring extra-dimensional physics and deepen our understanding of its potential implications for particle physics and cosmology.


\section*{Acknowledgment}

G.A. and R.N.C.F. acknowledge the financial support provided by Fundação Cearense de Apoio ao Desenvolvimento Científico e Tecnológico (FUNCAP) and the Conselho Nacional de Desenvolvimento Científico e Tecnológico (CNPq). We also express our gratitude to the referee for their insightful comments and suggestions, which have significantly improved the clarity and depth of this work. Their critical observations regarding the interpretation of field localization and the role of the ansatz have provided valuable directions for further research.  

Furthermore, we sincerely appreciate Alexandra Elbakyan and Sci-Hub's unwavering commitment to open science. By removing barriers to access, they have played a crucial role in democratizing knowledge and enabling researchers worldwide to pursue scientific discovery without restrictions.


\end{document}